**Boron and nitrogen isotope effects on hexagonal boron nitride properties**

E. Janzen,[1] H. Schutte,[1] J. Plo,[2] A. Rousseau,[2] T. Michel,[2] W. Desrat,[2] P. Valvin,[2] V. Jacques,[2] G. Cassabois,[2] B. Gil,[2] and J.H. Edgar[1*]

[1] Kansas State University, Tim Taylor Department of Chemical Engineering, 1005 Durland Hall, 1701A Platt St., Manhattan, KS 66506-5102

[2] Laboratoire Charles Coulomb, Université de Montpellier and CNRS, 34095 Montpellier, France

*Corresponding author. E-mail : edgarjh@ksu.edu

**Abstract**
The unique physical, mechanical, chemical, optical, and electronic properties of hexagonal boron nitride (hBN) make it a promising two-dimensional material for electronic, optoelectronic, nanophotonic, and quantum devices. Here we report on the changes in hBN's properties induced by isotopic purification in both boron and nitrogen. Previous studies on isotopically pure hBN have focused on purifying the boron isotope concentration in hBN from its natural concentration (approximately 20 at% $^{10}$B, 80 at% $^{11}$B) while using naturally abundant nitrogen (99.6 at% $^{14}$N, 0.4 at% $^{15}$N), *i.e.* almost pure $^{14}$N. In this study, we extend the class of isotopically-purified hBN crystals to $^{15}$N. Crystals in the four configurations, namely h$^{10}$B$^{14}$N, h$^{11}$B$^{14}$N, h$^{10}$B$^{15}$N, and h$^{11}$B$^{15}$N, were grown by the metal flux method using boron and nitrogen single isotope (>99%) enriched sources, with nickel plus chromium as the solvent. In-depth Raman and photoluminescence spectroscopies demonstrate the high quality of the monoisotopic hBN crystals with vibrational and optical properties of the $^{15}$N-purified crystals at the state of the art of currently available $^{14}$N-purified hBN. The growth of high-quality h$^{10}$B$^{14}$N, h$^{11}$B$^{14}$N, h$^{10}$B$^{15}$N, and h$^{11}$B$^{15}$N opens exciting perspectives for thermal conductivity control in heat management, as well as for advanced functionalities in quantum technologies.

**Introduction**

Hexagonal boron nitride (hBN) is the second most important two-dimensional material after graphite and its monolayer derivative, graphene. Although hBN and graphite have similar crystal structures and lattice parameters, their properties are quite different. Graphite is electrically conductive; hBN is an electrical insulator. Graphite is opaque; hBN is optically transparent from deep ultraviolet through the infrared. Graphite reacts with oxygen and air at low (~400°C) temperatures, while hBN is chemically inert. Thus, in many ways, the properties of hBN and graphite are complimentary.

The properties of crystalline hBN are advantageous for a wide range of nanophotonic [1], electronic [2,3], and quantum [4,5] devices. Because its optical properties are hyperbolic, it can compress infrared light to extremely small volumes, through the use of hyperbolic phonon polaritons. This is useful in nanophotonics as it increases light-matter interactions, and thereby chemical and biological sensing [6]. It also enables much higher image resolutions, hyperlensing [7], than is possible using the source free-space light. hBN analogue switches with high power handling and switching speeds have shown promise for high frequency devices [8], while hBN-based memristors integrated into CMOS devices for computation had superior endurance compared to more mature technologies [9]. hBN is appealing for quantum devices due its ability to host bright, stable, single photon emitters, some of which are magnetically addressable [10,11]. Boron nitride is also an excellent platform for near zero index photonics in the infrared [12]. Near-zero-index materials exhibit unique features in the area of light–matter interaction: when the values of the relative permittivity reach zero, at the operating frequency, the wavelength in these media is "stretched," and the phase of the signal is approximately uniform across this structure, which shapes the time oscillating waves to a static-like spatial distribution. An eclectic series of interesting and exciting phenomena can thus be initiated, including the levitation of electrically polarized nanoparticles in the vicinity of such material [13].

We have previously demonstrated that the properties of hBN can be varied by controlling its concentrations of boron isotopes. There are two stable boron isotopes, which are naturally distributed as approximately 20% $^{10}$B and 80% $^{11}$B. These isotopes have distinctly different properties. Their nuclear spins are 3/2 and 3 for $^{10}$B and $^{11}$B respectively. $^{10}$B has one of the largest neutron capture cross section of any isotope of any element, approximately 3855 barns for neutrons with velocities of 2200 m/s [14]. For the majority of element isotopes, the neutron capture cross section is on the order of one or less.

The hBN crystalline flakes enriched to nearly 100% $^{10}$B and 100% $^{11}$B (described as monoisotopic), as well as in variable proportions, were grown by precipitating from molten metal solutions at atmospheric pressure [15-19]. Typically, the solvents were nickel plus chromium [15], but hBN crystals were also grown using an iron plus chromium solvent [16]. The source materials necessary to synthesize this hBN were isotopically-enriched boron powder and ultra-high purity nitrogen gas with its natural isotope distribution (99.6% $^{14}$N, 0.4% $^{15}$N). The nitrogen was supplied by continuously flowing nitrogen gas over the solution during the entire process. A small concentration of hydrogen was added to the gas flow during crystal growth, to reduce and remove any oxides present. The molten solution was created by heating the components (contained in an alumina crucible) to 1550 °C, then slowly cool at 1 to 4 °C/h to precipitate the crystals.

The properties of boron isotope enriched hBN were characterized in several studies. Vuong et al [17] demonstrated that the energy band gap of hBN was slightly reduced or increased when enriched in $^{10}$B or $^{11}$B, respectively. The $E_{2g}^{high}$ peak position in the Raman spectra shifted up from 1,366 cm$^{-1}$ for hBN with the natural distribution of boron isotopes to 1,393 cm$^{-1}$ for h$^{10}$BN and down to 1,357 cm$^{-1}$ for h$^{11}$BN, [17-19]. The FWHM of the $E_{2g}^{high}$ peak at room temperature of the boron monoisotopic hBN was also reduced to 3.0 cm$^{-1}$ from 7.5 cm$^{-1}$ for natural hBN, due to less isotopic disorder [19]. The maximum peak width was predicted to occur with isotope concentrations of 65% $^{11}$B and 35% $^{10}$B [19]. The phonon lifetime for monoisotopic hBN is increased by a factor of 2.7 from h$^{nat}$BN [20]. Using scattering type near field optical microscopy (S-NOM) Giles et al [18] first showed the phonon-polariton propagation length is significantly increased in monoisotopic hBN. Pavlidis et al [21] measured the maximum hyperbolic phonon polariton lifetimes as 4.2 ps with propagation lengths of 25 µm, and an eight-fold increased figure of phonon propagating merit in comparison to h$^{nat}$BN.

The longer phonon lifetimes achieved by eliminating isotopic disorder increases the thermal conductivity of hBN. Lindsay and Broido [22], first predicted the thermal conductivity of hBN increase by eliminating isotopic disorder, ie for hBN with a single boron isotope. They also predicted the thermal conductivity of atomically thin layers of hBN would be much higher than bulk hBN [22]. Yuan et al [23] measured a maximum in-plane room temperature thermal conductivity of 585 W/(m·K) for bulk h$^{10}$BN and 550 W/(m·K) for h$^{11}$BN, compared to 408 W/(m·K) for h$^{nat}$BN. Mercado et al [24] reported a similar value of 630 W/(m·K) for h$^{10}$BN five layers thick. Subsequently, on a h$^{11}$BN monolayer, Cai et al [25,26] measured an in-plane thermal conductivity of 1009 W/(m·K), compared to 751 W/(m·K) for h$^{nat}$BN, as well as noting differences in the coefficients of thermal expansion with boron isotope. Hot carriers cool faster when generated in graphene in contact with higher thermal conductivity monoisotopic hBN than hBN with the natural distribution of boron isotopes [27].

Monoisotopic hBN has also proven advantageous in several applications. The longer hyperbolic phonon polariton propagation distance in isotopically enriched h$^{10}$BN enabled Li et al [28] to produce an anomalous, concave wave front on a h$^{10}$BN metasurface. Demonstrating hyperlensing, objects as small as 44 nm separated by distances of 25 nm were resolved via N-SOM using light with a free-space wavelength of 6.76 µm and h$^{11}$BN [7]. Vibrational strong coupling between phonon polaritons in hBN and organic molecules was reported by Bylinkin et al [29]. This has the potential to reduce the size and enhance the sensitivity of on-chip spectroscopes. Folland et al [30] changed the optical properties of h$^{11}$BN by placing it in contact with VO$_2$ which was changed from a dielectric to a metal by changing its temperature. Finally, isotopic purification with $^{10}$B leads to an improvement of the spin coherence properties of optically-active point defects in hBN, which are currently attracting a deep scientific interest for applications in quantum sensing and metrology [31].

In the present study, we demonstrate further control over the properties of hBN by changing the concentration of the nitrogen isotope in crystalline hBN. Natural nitrogen is predominantly $^{14}$N (99.6%) with a small amount of $^{15}$N (0.4%). Changing the nitrogen isotopes as well as the boron isotopes offers the possibility of shifting its Reststrahlen band for better matching to the infrared absorption of specific organic molecules, to achieve greater sensitivity [32]. $^{14}$N and $^{15}$N have different nuclear spins, 1 and ½ respectively, which may prove important in quantum applications. Unlike $^{14}$N, $^{15}$N reacts with high energy protons to form carbon and an alpha particle [33].

There have only been a few prior publications on the effects of $^{15}$N in boron nitride [32,34,35]. He *et al* [32] examined the impacts of changes with both boron and nitrogen isotopes on dielectric properties of cubic boron nitride. Jesche *et al* [34] studied the nuclear magnetic resonance spectra combination of all stable boron and nitrogen isotopes in both cubic boron nitride (cBN) and hBN containing the natural distributions of isotopes (*i.e.* no isotope enrichment). The present study is distinguished from these prior studies by focusing on high quality hBN crystals enriched with the nitrogen-15 isotope.

**Synthesis**
Previous work on atmospheric pressure solution growth of hBN with monoisotopic boron ($^{10}$B or $^{11}$B) and natural abundance nitrogen (99.6% $^{14}$N) used a continuous stream of nitrogen gas that both provides nitrogen for the formation of hBN and sweeps volatile impurities out of the system. However, the high cost of monoisotopic $^{15}$N gas would make continuous $^{15}$N$_2$ gas flow financially unsustainable. Therefore, for the synthesis of hBN with monoisotopic $^{15}$N, a new two-step process was developed to make more efficient use of nitrogen.

The first step, termed ingot formation, was done to purify the source materials and thoroughly mix them into an isotropic solution without any nitrogen. Isotopically pure (>99%) $^{10}$B or $^{11}$B powder was mixed with Ni and Cr powder in an alumina crucible in the mass ratio 12/12/1 Ni/Cr/B, respectively. This was then placed in a horizontal tube furnace, evacuated of air, and purged with a mixture of 93/7 argon/hydrogen flowing continuously at 270 sccm and 850 torr. The furnace was heated to 1550°C, held there for 24 hours, and cooled back down as illustrated by the dashed line in Figure 1.

Separating this step from crystal growth has a couple notable benefits. First, it enables a precise and reproducible concentration of components to be used by ensuring all the source materials melt together into a cohesive flux. If nitrogen is used in this step, undissolved powder usually remains separated from the solidified flux, which changes the concentration of the flux by an unknown amount. Furthermore, if nitrogen were used in this step, it would need to be isotopically pure $^{15}$N$_2$ gas to avoid producing hBN that has a mixture of $^{14}$N and $^{15}$N. Second, hydrogen diluted with argon for safety is included to react with residual oxygen in the gas phase and to reduce oxides in the source materials to produce water vapor. Since no expensive $^{15}$N$_2$ gas is used in this step, the gas flow can be continuous, sweeping the oxygen impurities from the system, thus purifying the flux for crystal growth.

In the crystal growth step, the ingot is heated and slowly cooled in the presence of $^{15}N_2$ gas to precipitate hBN enriched with $^{15}N$. The ingot produced in the first step is placed in the same horizontal tube furnace purged of air using argon, then backfilled with a mixture isotopically pure $^{15}N_2$ gas and hydrogen. To limit the use of $^{15}N$, the tube was sealed off once the pressure rose to atmospheric pressure and hydrogen was included to react with any residual oxygen. Then the furnace was heated to 1550°C, held there for 24 hours, and slowly cooled at 1°C/hr to 1500°C to precipitate hBN single crystals as illustrated by the solid line in Figure 1. Once the system has cooled, the alumina boat contains a solidified metal ingot with a layer of hBN crystals covering its top surface, as shown in Figure 2. Mechanical exfoliation of the hBN from the ingot produces free-standing hBN flakes like the one shown in Figure 3, which were used in Raman and photoluminescence measurements.

Since isotopically pure boron ($^{10}B$ or $^{11}B$) and isotopically pure $^{15}N$ are the only boron and nitrogen sources present in the system during the process, any hBN that forms will have these isotopes. Thus, hBN can now be produced with any combination of boron and nitrogen isotopes: h$^{10}B^{14}N$, h$^{11}B^{14}N$, h$^{10}B^{15}N$, or h$^{11}B^{15}N$. This process also has the flexibility to produce hBN with precise mixtures of isotopes by controlling how much $^{10}B$ and $^{11}B$ powder is added in the first step and the ratio of $^{14}N_2$ and $^{15}N_2$ gas introduced to the system, though that was not attempted in this study.

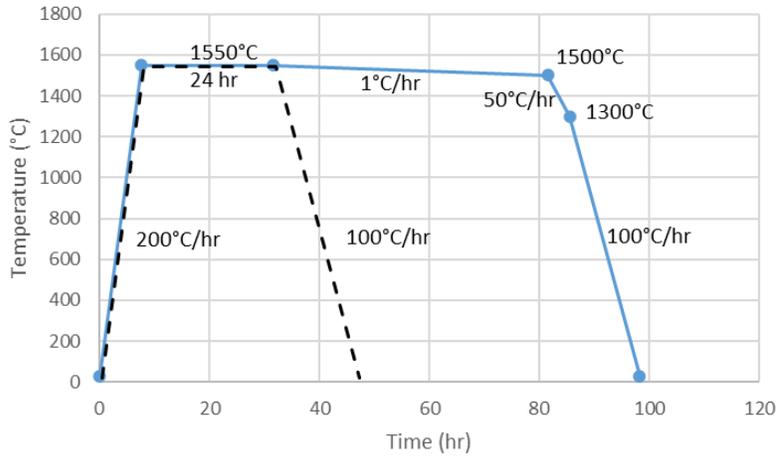

Figure 1 – Temperature profile of two-part process used to produce isotopically-controlled hBN crystals. In the first part (dashed line), the system is heated and dwelled at 1550°C under an Ar/$H_2$ atmosphere to form a homogenous ingot. In the second part (solid line), the system is heated to 1550°C and slowly cooled under a $^{15}N_2$/$H_2$ atmosphere to precipitate hB$^{15}$N crystals.

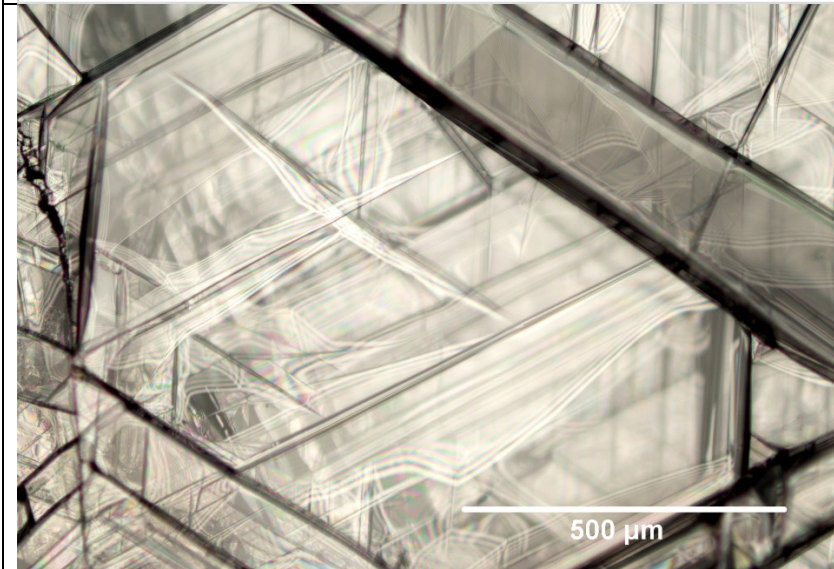

Figure 2 – Image of the surface of the Ni-Cr-B ingot after an experiment with hB$^{15}$N single crystals on its surface.

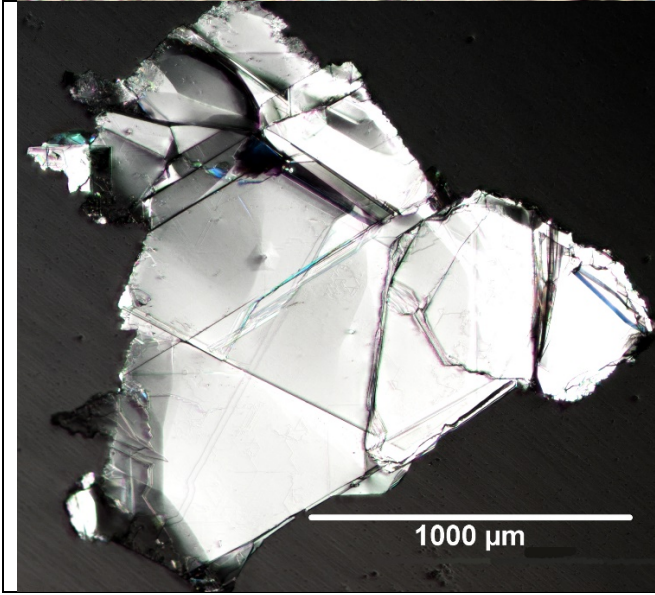

Figure 3 – Micrograph of a h$^{10}$B$^{15}$N single crystal mechanically exfoliated from a solidified ingot.

Results

Raman spectroscopy was carried out at room temperature. The Raman mode at ~1360 cm$^{-1}$ was measured with a Renishaw Invia spectrometer (2400 lines/mm) and the mode at ~50 cm$^{-1}$ was measured using a Horiba T64000 (1800 lines/mm) in the subtractive mode for which we used the oversampling method implemented in Ref.[17]. For the low- (resp. high-) frequency measurement of the $E_{2g}$ Raman phonon, a 660 (resp. 532) nm laser was focused under a 50x objective (N.A. 0.5) with the power at the sample kept below 300 µW to avoid heating effects. The high-energy Raman mode ($E_{2g}^{high}$) was studied in monoisotopic h$^{10}$B$^{14}$N, h$^{11}$B$^{14}$N, h$^{10}$B$^{15}$N, and h$^{11}$B$^{15}$N crystals, together with a reference h$^{nat}$BN crystal with natural abundance of the boron and nitrogen isotopes [Figure 4]. Compared to previous Raman results [17], in the data set of Figure 4, the new h$^{10}$B$^{15}$N and h$^{11}$B$^{15}$N configurations incorporate Raman lines that are roughly the mirror images of h$^{11}$B$^{14}$N and h$^{10}$B$^{14}$N with respect to h$^{nat}$BN. This results in a progressive red-shift of the $E_{2g}^{high}$ mode as a function of the reduced mass in the unit cell, from ~1393 cm$^{-1}$ in h$^{10}$B$^{14}$N to ~1335 cm$^{-1}$ in h$^{11}$B$^{15}$N [Figure 4b]. The experimental blue-shifts (red-shifts) of +27.8 and +9.7 cm$^{-1}$ (-8.3 and -27.6 cm$^{-1}$) in h$^{10}$B$^{14}$N and h$^{10}$B$^{15}$N (h$^{11}$B$^{14}$N and h$^{11}$B$^{15}$N) with respect to h$^{nat}$BN are in excellent agreement with the +30.9 and +11.1 cm$^{-1}$ (-7.0 and -27.4 cm$^{-1}$) values computed by means of DFT calculations within the QuantumEspresso package, respectively. Remarkably, all monoisotopic crystals display narrow Raman lines with a width ~4 cm$^{-1}$, significantly smaller than the ~8 cm$^{-1}$ width in h$^{nat}$BN. This effect stems from the suppression of isotopic disorder in monoisotopic crystals [17], and direct evidence of the equally-high crystalline quality of the h$^{10}$B$^{14}$N, h$^{11}$B$^{14}$N, h$^{10}$B$^{15}$N, and h$^{11}$B$^{15}$N crystals.

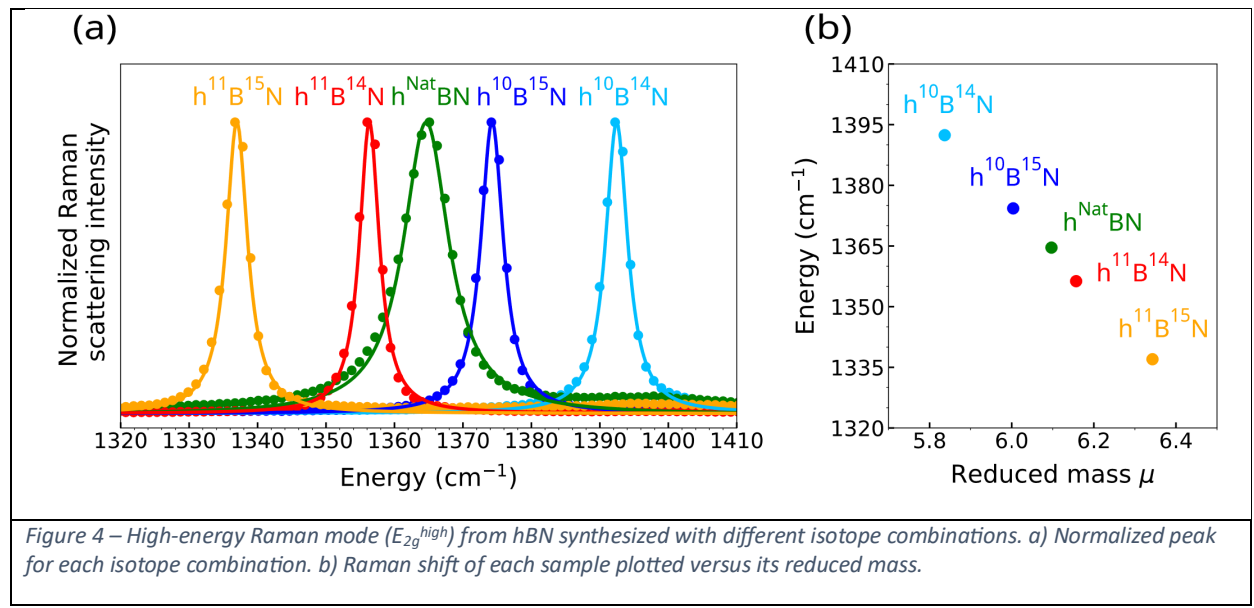

*Figure 4 – High-energy Raman mode ($E_{2g}^{high}$) from hBN synthesized with different isotope combinations. a) Normalized peak for each isotope combination. b) Raman shift of each sample plotted versus its reduced mass.*

For the low-energy Raman mode ($E_{2g}^{low}$), the behavior is different [Figure 5]. This mode corresponds to the shear rigid motion between adjacent layers, with a frequency around 50 cm$^{-1}$ [Figure 5a]. In contrast to the smooth variations of the $E_{2g}^{high}$ frequency as a function of the reduced mass and the related mirror symmetry of the Raman lines with respect to h$^{nat}$BN [Figure 4], the $E_{2g}^{low}$ mode does not shift monotonously with the isotopic composition, as seen in Figure 5b. First, the $E_{2g}^{low}$ frequency recorded in the reference $h^{nat}BN$ crystal is no longer the average value of the ones in $h^{10}B^{14}N$ and $h^{11}B^{15}N$. Second, for $h^{10}B^{15}N$, a red-shift of the $E_{2g}^{low}$ mode is recorded, instead of the blue-shift for the high-frequency Raman line [Figure 4], which is naturally explained by the variation of the reduced mass. Strikingly, the $E_{2g}^{low}$ frequency turns out to be approximately the same for $h^{10}B^{15}N$ and $h^{11}B^{14}N$. This complex phenomenon reveals the non-trivial variations of the interlayer shear mode frequency with the isotopic composition, and thus the subtle impact of the isotopic content on the weak van der Waals interaction between adjacent layers. Although the $E_{2g}^{high}$ mode grossly varies linearly versus the reduced mass [Figure 4 (b)], the $E_{2g}^{low}$ one does not and Figure 5(b) indicates a parallelogram-like locus with four corners defined by the energies of the $^{10}B^{14}N$, $^{10}B^{15}N$, $^{11}B^{15}N$ and $^{11}B^{14}N$ samples. This locus is the experimental evidence of the delicate dependences of the phonon modes with isotopic composition. Note for the sake of the completeness that the $^{Nat}BN$ value sits on the line connecting $^{10}B^{14}N$ to $^{11}B^{14}N$. The deviation is linked to alloy disorder in the boron sublattice [36].

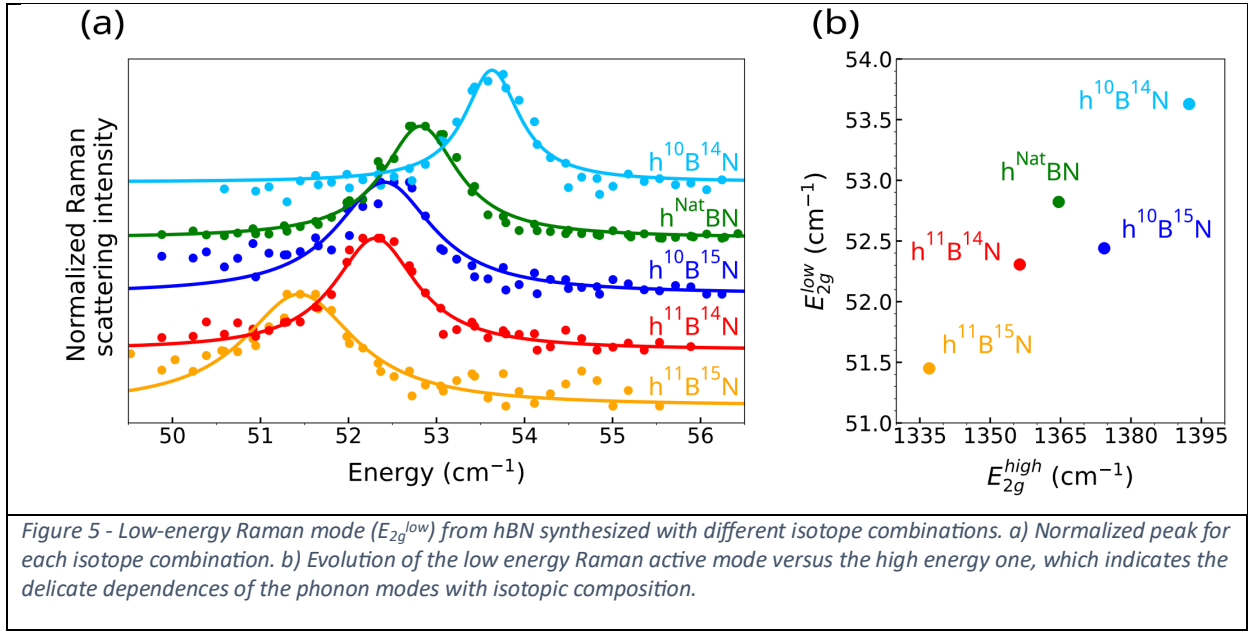

*Figure 5 - Low-energy Raman mode ($E_{2g}^{low}$) from hBN synthesized with different isotope combinations. a) Normalized peak for each isotope combination. b) Evolution of the low energy Raman active mode versus the high energy one, which indicates the delicate dependences of the phonon modes with isotopic composition.*

These results are quantitatively interpreted by our DFT calculations of the dispersion relations of the phonon modes in these samples that are offered in Figure 6. Four specific regions have been zoomed in four inserts dedicated to the $E_{2g}^{low}$ and the $E_{2g}^{high}$ normal vibration modes. The plots on the left-hand side correspond to the center of the Brillouin Zone while the plots on the right-hand side represent the dispersions relations of the (TA,LA) and (TO,LO) couples at the T point of the Brillouin Zone. The values of the energies of the latter are useful for an accurate determination of the values of the indirect excitons versus isotropic purification as we shall discuss it later in this manuscript. They also account for the asymmetric shifts of the $E_{2g}^{low}$ mode in $h^{10}B^{14}N$ and $h^{11}B^{15}N$ with theoretical estimations of +0.9 and -1.3 cm$^{-1}$ with respect to $h^{nat}BN$, in fair agreement with the experimental values of +0.8 and -1.4 cm$^{-1}$. Moreover, DFT calculations predict the same frequency within an accuracy of 0.1 cm$^{-1}$ for the $E_{2g}^{low}$ mode in $h^{10}B^{15}N$ and $h^{11}B^{14}N$, with a red-shift of -0.2 cm$^{-1}$ with respect to $h^{nat}BN$, thus providing a consistent interpretation of our experimental data where the measured red-shift is ~ -0.45 cm$^{-1}$.

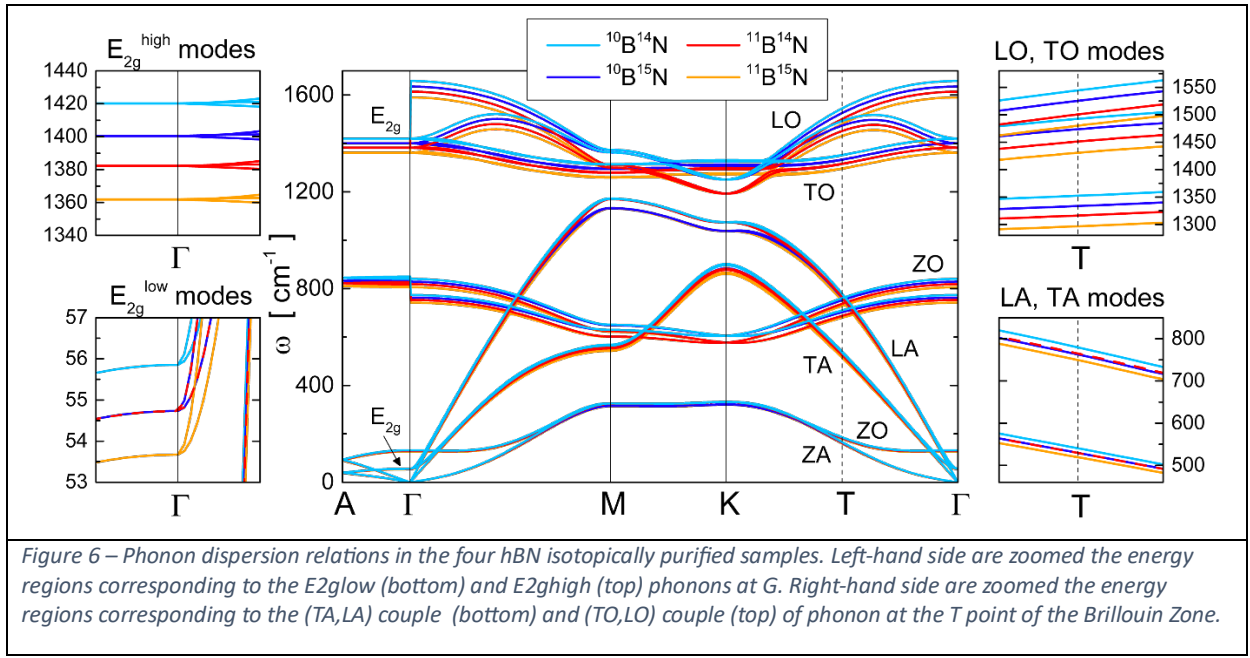

*Figure 6 – Phonon dispersion relations in the four hBN isotopically purified samples. Left-hand side are zoomed the energy regions corresponding to the E2glow (bottom) and E2ghigh (top) phonons at G. Right-hand side are zoomed the energy regions corresponding to the (TA,LA) couple (bottom) and (TO,LO) couple (top) of phonon at the T point of the Brillouin Zone.*

We performed photoluminescence (PL) spectroscopy in the deep-ultraviolet to further characterize the monoisotopic hBN crystals and assess their optical properties. By means of the experimental setup described previously [17], we recorded the intrinsic emission at the hBN band-edge at wavelengths ~200 nm for samples at cryogenic temperatures (T=8K). For all $h^{10}B^{14}N$, $h^{11}B^{14}N$, $h^{10}B^{15}N$, and $h^{11}B^{15}N$ crystals, and the reference $h^{nat}BN$ one, intense emission with narrow PL lines is recorded, attesting again to high crystalline quality of the samples [Figure7a]. As with the high-energy Raman mode [Figure 4], the PL spectrum varies smoothly with the isotopic content, with a progressive blue-shift as a function of the reduced mass [Figure7a]. Similar to the $E_{2g}^{high}$ Raman-mode, the new $h^{10}B^{15}N$ and $h^{11}B^{15}N$ configurations appear as "mirror images" of $h^{11}B^{14}N$ and $h^{10}B^{14}N$ with respect to $h^{nat}BN$ in the PL data set of [Figure 7]. The shift of the PL spectra with isotopic composition stems from the indirect nature of the bandgap in bulk hBN, where the intrinsic emission results from phonon-assisted recombination [17,36-39]. Since all PL lines are phonon replicas in Figure 7a, their spectral position directly depends on the energy of the specific phonon mode involved in the recombination process [17], hence on the isotopic composition of the hBN crystal. The lines at ~5.77 eV are the most detuned from the indirect bandgap because they correspond to the emission of LO phonons, of highest energy in the phonon band structure [36,40]. As a consequence, the isotopic shift is also maximal for LO phonons, thus explaining the increasing isotopic shift of the PL lines on reducing the detection energy in the PL spectra of Figure7a. We performed a quantitative interpretation of the PL spectra following the procedure previously employed in Ref.[17]. By plotting the energy of the different phonon replicas as a function of the calculated values of the phonon energy at the T point of the Brillouin zone, we extracted the energy of the fundamental indirect exciton for each monoisotopic hBN crystal [Figure7b]. Such an analysis reveals the few-meV variations of hBN bandgap with the isotopic composition, which come from the zero-point renormalization of the bandgap energy [17]. The possibility to investigate such a fine effect is again a signature of the high crystalline quality of all h$^{10}$B$^{14}$N, h$^{11}$B$^{14}$N, h$^{10}$B$^{15}$N, and h$^{11}$B$^{15}$N crystals.

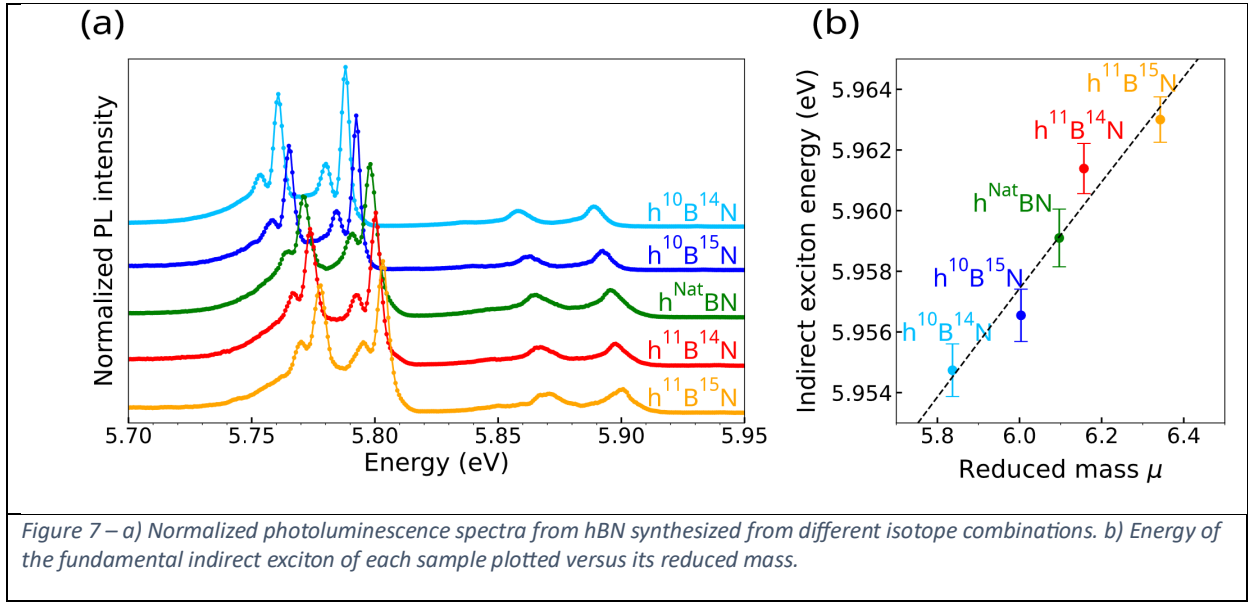

Figure 7 – a) Normalized photoluminescence spectra from hBN synthesized from different isotope combinations. b) Energy of the fundamental indirect exciton of each sample plotted versus its reduced mass.

Conclusions

In this work, we have developed a new 2-step process for the crystal growth of hBN with monoisotopic $^{15}$N that efficiently uses $^{15}$N$_2$ gas and can produce hBN crystals with any isotope combination: h$^{10}$B$^{14}$N, h$^{11}$B$^{14}$N, h$^{10}$B$^{15}$N, or h$^{11}$B$^{15}$N. The first step consists of the formation of an homogenous ingot from isotopically pure boron powder ($^{10}$B or $^{11}$B) mixed with Ni and Cr powder while removing oxygen impurities. The second step consists of the growth of hBN crystals from the ingot using a sealed tube containing $^{15}$N$_2$ gas and hydrogen. This step produces hBN crystals with a comparable size and clarity to those achieved by previous methods.

Raman spectroscopy reveals the achievement of $^{15}$N-purified hBN crystals of the same high-quality as for $^{14}$N, as demonstrated by the experimental ~4 cm$^{-1}$-width of the high-energy Raman line. On the one hand, the frequency of the high-energy Raman mode progressively red-shifts with the reduced mass in the unit cell, from ~1393 cm$^{-1}$ in h$^{10}$B$^{14}$N to ~1335 cm$^{-1}$ in h$^{11}$B$^{15}$N. On the other hand, the behavior of the low-energy Raman line corresponding to the interlayer shear mode is non-monotonous, with strikingly the same approximate frequency in h$^{11}$B$^{14}$N and h$^{10}$B$^{15}$N crystals. Our results agree with our *ab-initio* calculations of the vibrational properties as a function of the isotopic composition.

The high-quality of the four isotopically-purified hBN crystals is further proven by photoluminescence (PL) spectroscopy excited in the UV-C spectral range, above the hBN bandgap. Intense and narrow PL lines are recorded, with a smooth blue-shift of the PL spectrum as a function of the reduced mass, as expected for indirect-gap bulk hBN where the intrinsic band-edge emission stems from phonon-assisted recombination. Quantitative analysis of the PL spectra allows to extract the variations of the indirect bandgap with the isotopic composition.

The work expands the range of properties that can be achieved by varying the isotopic composition of hBN. This will be a major benefit for tailoring the properties to optimize the performance of electronic, optoelectronic, nanophotonic, and quantum devices based on hBN.

**Acknowledgements**
Office of Naval Research award N00014-22-1-2582 provided support for the hexagonal boron nitride crystal growth. Optical spectroscopy was financially supported by the BONASPES project (ANR-19-CE30-0007), the ZEOLIGHT project (ANR-19-CE08-0016), and the HETERO-BNC project (ANR-20-CE09-0014-02).

**Conflicts of Interest**
None of the authors have an financial/commercial Conflicts of Interest with this work.